\documentclass{article}
\usepackage{spconf,amsmath,graphicx}
\usepackage{tabularx}
\usepackage{booktabs}
\usepackage{multirow}
\usepackage{cleveref}
\usepackage{flushend}
\usepackage{placeins}
\usepackage{url}
\usepackage{xspace}
\usepackage{siunitx}
\usepackage[shortcuts]{extdash}
\newcommand{\fb}[0]{FluencyBank\xspace}
\newcommand{\sep}[0]{SEP\=/28k\xspace}
\newcommand{\sepE}[0]{SEP\=/28k\=/E\xspace}
\newcommand{\ksof}[0]{KSoF\xspace}

\title{Dysfluencies Seldom Come Alone -- Detection as a Multi-label problem}
\name{Sebastian P. Bayerl$^{\star}$,
Dominik Wagner$^{\star}$, 
F. Hönig$^{\ddagger}$,
T. Bocklet$^{\star}$$^{\diamond}$, 
E. Nöth$^{\dagger}$,
Korbinian Riedhammer$^{\star}$}

\address{
$^{\star}$ Technische Hochschule Nürnberg Georg Simon Ohm, Germany \\
$^{\ddagger}$KST Institut GmbH, $^{\diamond}$Intel Labs\\
$^{\dagger}$ Friedrich-Alexander-Universität Erlangen-Nürnberg, Germany
}

\begin{document}
\maketitle
\begin{abstract}
Specially adapted speech recognition models are necessary to handle stuttered speech. 
For these to be used in a targeted manner, stuttered speech must be reliably detected.
% Voice technology must be adapted to be able to handle atypical speech which in a first step needs to be detected reliably. 
% To be able topatterns need to be detected to apply models that can handle such speech patterns.
% Atypical speech patterns need to be detected to apply models that can handle such speech patterns.
Recent works have treated stuttering as a multi-class classification problem or viewed detecting each dysfluency type as an isolated task; that does 
not capture the nature of stuttering, where one dysfluency seldom comes alone, i.e., co-occurs with others.
% dysfluencies.
This work explores an approach based on a modified \textit{wav2vec 2.0} system for end-to-end stuttering detection and classification as a multi-label problem. 
The method is evaluated on combinations of three datasets containing English and German stuttered speech, yielding state-of-the-art results for stuttering detection on the {SEP-28k-Extended} dataset.
Experimental results provide evidence for the transferability of features and the generalizability of the method across datasets and languages.

\end{abstract}
\begin{keywords}
stuttering, dysfluency detection, dysfluency, cross-dataset, pathological speech
\end{keywords}

\vspace{-2mm}
\section{Introduction}\label{sec:intro}
\vspace{-2mm}
Dysfluency means abnormality of fluency, which includes, but is not limited to, stuttering \cite{wingate_FluencyDisfluencyDysfluency_1984}.
Stuttering is a complex fluency disorder that can be identified by its core symptoms; repetitions of words, syllables, and sounds, prolongations, and blocks while speaking \cite{lickley_DisfluencyTypicalStuttered_2017}.
Therapeutic options, including modifying one's speech, have been proposed to alleviate the symptoms and improve fluency.
% Stuttering can be debilitating to a person's ability to communicate, which also extends to the use of voice technology which works worse for people with atypical speech patterns.
Stuttering can be debilitating to a person's ability to communicate, which also extends to voice technology, making it necessary to detect atypical speech reliably and apply custom models.
% Atypical speech patterns are a challenge for automatic speech recognition (ASR) systems. 
% voice technology?
% Making it necessary to reliably detect atypical speech in order to apply custom models. 
% Speech therapy that teaches speech modifications does also not lead to a normalized speech pattern, but to an effective one. 
% To this ends methods to reliably detect and classify atypical speech patterns are needed. 

Recent work on machine learning for stuttering has concerned itself with stuttering detection \cite{bayerl_DetectingDysfluenciesStuttering_2022a, harvill_FramelevelStutterDetection_2022,kourkounakis_DetectingMultipleSpeech_2020} and stuttering classification \cite{grosz_Wav2vec2basedParalinguisticSystems_2022,sheikh_EndtoEndSelfSupervisedLearning_2022,montacie_AudioFeaturesWav2Vec_2022,you_MaskedModelingbasedAudio_2022}. 
Grosz et al. used pre-trained German wav2vec 2.0 (W2V2) models and combined them with other classifiers in an ensemble approach \cite{grosz_Wav2vec2basedParalinguisticSystems_2022}.
The authors of \cite{montacie_AudioFeaturesWav2Vec_2022} also used W2V2 features, treating them like low-level descriptors, and computed several functionals on the features, similarly to the openSMILE approach \cite{eyben_OpensmileMunichVersatile_2010}.

In \cite{kourkounakis_DetectingMultipleSpeech_2020}, the authors describe an approach employing long short-term memory (LSTM) networks with a residual neural network (ResNet) backend to detect stuttering in the UCLASS corpus \cite{howell_UniversityCollegeLondon_2009,kourkounakis_DetectingMultipleSpeech_2020}.
Lea et al. applied multi-task learning with LSTMs in their dysfluencies detection approach \cite{lea_SEP28kDatasetStuttering_2021}.
Dysfluency detection systems trained on one dataset generalized to another, with data quantity being the deciding factor for detection performance \cite{lea_SEP28kDatasetStuttering_2021}. 
% but did not explore training on multiple datasets or fine-tuning \cite{lea_SEP28kDatasetStuttering_2021}.
Recent work by \cite{bayerl_DetectingDysfluenciesStuttering_2022a} could show that W2V2 features can be fine-tuned for dysfluency detection using stuttering data from other datasets. 
The features extracted from models fine-tuned with stuttering data were transferable from English to German for all dysfluency types but word repetitions.
% W2V2 models fine-tuned on English stuttering data yield features that improve dysfluency detection systems on other stuttering datasets \cite{bayerl_DetectingDysfluenciesStuttering_2022}.  
Both works did not explore the effect of multi-dataset training on their detection systems.  
% Multi-dataset training of pathological speech systems remains under-explored.

The \sep dataset, and the relabeled \fb dataset, are fairly new resources.
No suggested evaluation split makes it hard to reproduce and compare results to the baseline systems \cite{lea_SEP28kDatasetStuttering_2021}.
Several researchers have used it to evaluate their methods but have failed to publish their exact splits or have filtered out examples that substantially impact evaluation results.
Sheikh et al. removed a class by combining word- and sound repetitions and filtered out clips with an additional non-stuttering label, e.g., natural pause or background music \cite{lea_SEP28kDatasetStuttering_2021}, i.e., removing difficult examples.
They randomly split the remaining data into a train, development, and test set (80/10/10\%), irrespective of the speakers \cite{sheikh_RobustStutteringDetection_2022}. 
% They removed one class by treating word- and sound repetitions as one, which are acoustically different and can be recognized using shorter or longer context.
% A major difficulty in the field is reproducibility of results due to the use of inconsistent labeling and evaluation strategies. 
% The same as the case for the \sep, \fb, and the \ksof~ dataset. 
% Several authors have published results on \sep, but failed to publish their data partition or did only use a subset of the data.  
Other work filtered out all non-unanimously labeled clips from the training and test data, aiming for easy samples \cite{mohapatra_SpeechDisfluencyDetection_2022}.
The work by \cite{jouaiti_DysfluencyClassificationStuttered_2022} left out the block class and did use random splitting irrespective of the speaker, which is known to lead to optimistic results, as the authors of \cite{bayerl_InfluenceDatasetPartitioning_2022} could show.
Furthermore, they explored the influence of dataset partitioning, created non-speaker overlapping splits of \sep and provided baseline experiments and evidence for the large variance of detection results due to dataset partitioning \cite{bayerl_InfluenceDatasetPartitioning_2022}.
% https://dl.acm.org/doi/pdf/10.1145/3539490.3539601 filter out ambigious data, statement: quantity less important than quality, is wrong, as no real eval scenario, wehn only non ambigious, whcih is the nature of stuttering, give one more read, the eval on FB is unclear, paper quality sub par

% At the same time,  stuttering detection and classification are often treated as two isolated problems. 
% Speech samples do either have a unique single label \cite{schuller_ACMMultimedia2022_2022} for an audio clip or treating each dysfluency as their own detection task \cite{bayerl_DetectingDysfluenciesStuttering_2022a, kourkounakis_DetectingMultipleSpeech_2020}. 

% All these works contribute to a better understanding of dysfluency detection but missing dataset splits makes reproducing these works difficult. 
% the results are difficult to reproduce and sometimes overlook the fact that one dysfluency seldom comes alone.
Apart from problems with reproducibility, most works ignore that one dysfluency seldom comes alone, i.e., dysfluency patterns co-occur. 
In {SEP\=/28k\=/Extended (\sepE)}, \fb, and \ksof, about 30\%, 36\%, and 21\% of clips were labeled with more than one dysfluency type -- making stuttering detection and classification  multi-label problems.
% \sepE, \fb, and \ksof, containing about 30\%, 36\%, and 21\% clips that were labeled with more than one dysfluency type -- making the problem a multi-label detection and classification problem. 
% the imbalance in the dataset w.r.t. the speakers, making it vulnerable to overoptimistic, speaker dependent models \cite{bayerl_InfluenceDatasetPartitioning_2022}.

% This paper explores dysfluency detection and classification as a multi-label problem, contributing a new W2V2-based end-to-end (E2E) method to detect and classify stuttered speech.
% The method is evaluated on three datasets, yielding state-of-the-art detection performance on the \sepE dataset split. 
% Furthermore, we provide conclusive evidence for the generalizability of the method by exploring multi-dataset and multi-language training for dysfluency detection.

This paper explores dysfluency detection and classification as a multi-label problem, contributing a new W2V2-based end-to-end (E2E) method to detect and classify stuttered speech and evaluating it on three datasets, yielding state-of-the-art detection performance on the \sepE dataset split. 
Furthermore, we provide conclusive evidence for the generalizability of the method by exploring multi-dataset and multi-language training for dysfluency detection.
\vspace{-4mm}
\section{Data}\label{sc:data}
\vspace{-3mm}
% \subsection{SEP-28k-E, FluencyBank, and \ksof}
In our experiments, we use three corpora containing 3-second long clips with stuttered speech; SEP-28k-Extended, FluencyBank, and the Kassel State of Fluency (\ksof) dataset \cite{bayerl_InfluenceDatasetPartitioning_2022,lea_SEP28kDatasetStuttering_2021,bayerl_KSoFKasselState_2022}. 
The SEP-28k-E contains English stuttered speech extracted from podcasts and is based on the SEP-28k corpus, extending it with speaker labels and a speaker-exclusive Train-Dev-Test split.\footnote{Online: \protect\url{https://tinyurl.com/yck9fmfv}}
% \footnote{Available: \protect\url{https://github.com/th-nuernberg/ml-stuttering-events-dataset-extended}}

The \sep corpus contains an additional 4144 English clips extracted from the interview part of the adults who stutter dataset of the \fb corpus \cite{bernsteinratner_FluencyBankNew_2018} that were labeled using the same protocol.
For evaluation purposes, we use the split defined by \cite{bayerl_DetectingDysfluenciesStuttering_2022a}.\footnote{\protect Online: \url{https://tinyurl.com/24vm6dec}}
All three datasets were labeled similarly with five types of dysfluencies; blocks, prolongations, sound repetitions, word repetitions, and interjections.
\ksof is a German dataset whose clips were additionally labeled with speech modifications, marking a clip as containing a person using a speech technique known as fluency shaping.
This is a technique persons who stutter (PWS) learn in stuttering therapy to help them overcome their stuttering \cite{bayerl_KSoFKasselState_2022}.
The label distribution of the complete dataset and the respective test partition can be found in \Cref{tab:distribution}.

% \vspace{-4mm}
% Multi-lingual and multi-dataset training
% \sep and \fb are almost balanced w.r.t. the labels of each core-symptoms.
% Whereas the \ksof~ dataset has only few labeled word repetitions and many blocks and compared to the other datasets fluent labels are underrepresented, as people int therapy try avoid stuttering by using modified speech (see \cref{tab:distribution}) \cite{lea_SEP28kDatasetStuttering_2021,bayerl_KSoFKasselState_2022}. 

To evaluate the effect of multi-lingual and -dataset training, we define three combinations of datasets.
% in addition to the \ksof, \fb, and \sepE splits. 
ALL-EN is a combination of \fb and \sepE, Multilingual-Small (Multi-S) is a combination of \ksof and \fb, and Multilingual (Multi) consists of \ksof, \fb, and \sepE. 
Systems training with Multi-S, Multi, and \ksof predict seven classes (Modified (Mod), Blocks (Bl), Interjections (Int), Prolongation (Pro), Sound Repetitions (Snd), Word Repetitions (Wd), No Dysfluencies (No-Df)).
The \fb, \sepE, and ALL-EN splits are trained to detect six classes (Bl, Int, Pro, Snd, Wd, No-Df).
% Besides   use these splits to evaluate cross-dataset and cross-language performance.  
\begin{table}[tb]
    % \centering
    \vspace{-2mm}
    \caption{Label distribution in \% of \sepE (28k-E), \fb (FB), and \ksof. The suffix '-T' indicates the test-set label distribution.}\label{tab:distribution}
\scalebox{0.87}{
\begin{tabular}{l|c|c|c|c|c|c}
    \toprule
\textbf{Label} & \textbf{28k-E} & \textbf{28k-E-T} & \textbf{FB} & \textbf{FB-T} & \textbf{\ksof} & \textbf{\ksof-T} \\
% \textbf{Label} & \textbf{\sepE} & \textbf{\sepE-T} & \textbf{FB} & \textbf{FB-T} & \textbf{\ksof} & \textbf{\ksof-T} \\
                      
\midrule
Bl          &       12.0   &  12.0  & 10.3  &   8.4  &   20.7  &  18.0      \\  
Int         &       21.2   &  19.5  & 27.3 &  31.0  &   13.0  &  20.2      \\         
Pro         &       10.0   &  10.1  &  8.1 &  10.3  &   12.0  &  17.3      \\ 
Snd         &        8.3   &   6.7  & 13.3 &  11.5  &   14.8  &  11.4      \\ 
Wd          &        9.8   &  10.5  & 10.4 &   9.65 &   3.9   &  3.7      \\    
Mod         &        -     &   -   &  -  &    -  &   24.4  &  18.1      \\  
No          &       56.9   &  58.2  & 54.1 &  56.9  &   24.8  &  29.4      \\   
\midrule
$\Sigma$    &       28177  &  6562  & 4144 &   785  &  5597   &   1253          \\   
\bottomrule
\end{tabular}
}
\end{table}

\vspace{-3mm}
\section{Method}\label{sec:method}

\begin{figure}[!htb]
    \centering
    \includegraphics[width=1.0\linewidth]{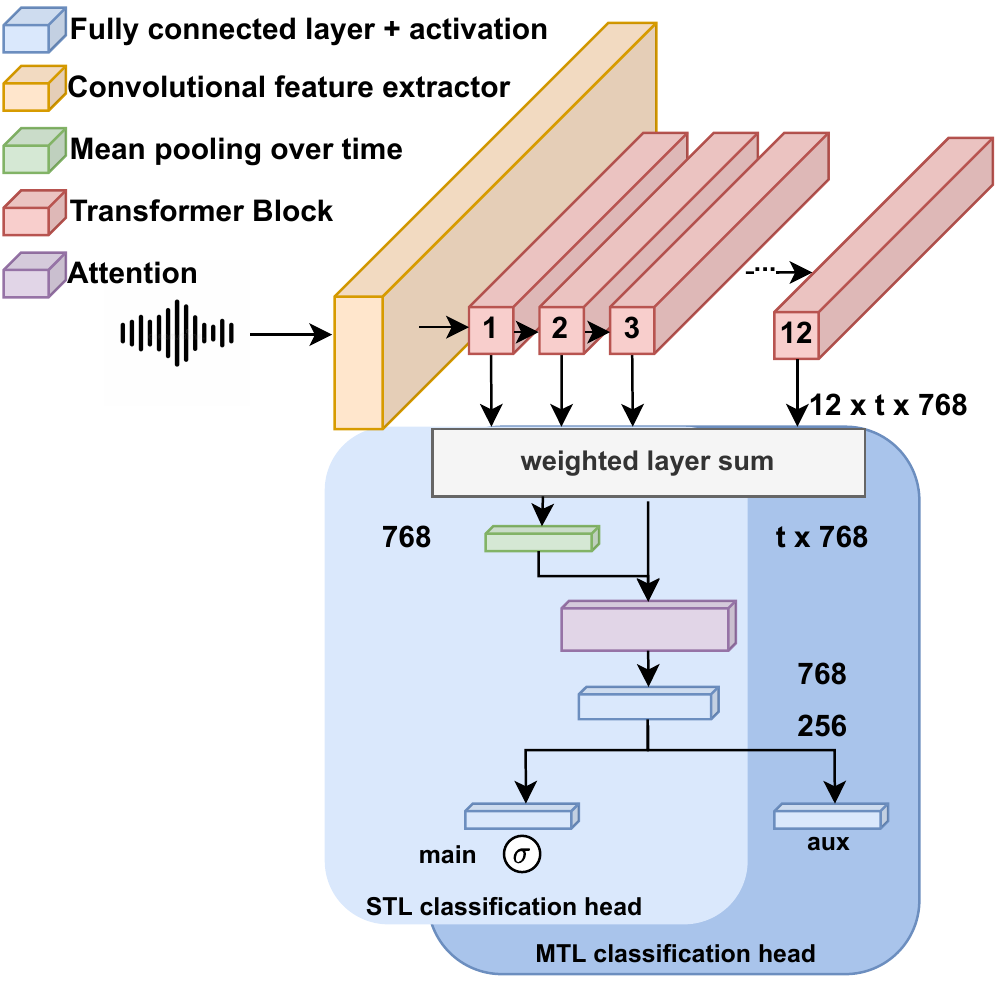}
    \caption{Schematic overview of the wav2vec 2.0 model with weighted layer sum ($\mathbf{WLS}$) and modified attention-based single and multi-task (STL, MTL) classification head.   
    }
  \label{fig:architecture}
  \vspace{-5mm}
\end{figure}

\subsection{wav2vec 2.0}\label{ss:w2v2}

The base W2V2 model consists of a convolutional feature extractor at the beginning of the model that takes in the waveform, followed by 12 transformer encoder blocks.  
The model encodes 20ms of audio into 768-dimensional feature vectors after each transformer block, yielding 12 x $t$ x 768 hidden representations \cite{baevski_Wav2vecFrameworkSelfSupervised_2020}. 
W2V2 features perform well in dysfluency detection \cite{bayerl_DetectingDysfluenciesStuttering_2022a} and other speech tasks, such as automatic speech recognition (ASR), and mispronunciation detection \cite{baevski_Wav2vecFrameworkSelfSupervised_2020,xu_ExploreWav2vecMispronunciation_2021}.

The model architecture was adapted from the W2V2 for sequence classification implementation from HuggingFace \cite{wolf_TransformersStateoftheArtNatural_2020}. 
The adaptation differs mainly in three aspects; Instead of using the outputs of the last hidden layer, all hidden states ($\mathbf{h_{l}}$) are multiplied by a trainable weight per layer ($w_{l}$), weighting its contribution to the final weighted layer sum ($\mathbf{WLS}$), resulting in a representation of $t \times 768$. 
\begin{equation}
\vspace{-2mm}
    \mathbf{WLS} = \sum_{l=1}^{12} w_{l} \mathbf{h_{l}}
\vspace{-1mm}
\end{equation}
Instead of using mean pooling along the time dimension, an attention mechanism takes the mean of all feature vectors along the time dimension as the query input $Q$, with keys $K$ and values $V$ being the $\mathbf{WLS}$.  % of all hidden-layer outputs.
The third change is a second output branch used for multi-task learning, analogous to the one described in \cite{bayerl_DetectingDysfluenciesStuttering_2022a}. 
A sigmoid function ($\sigma$) is applied to the main task outputs predicting the probability of each class for an audio clip.
% A sigmoid activation function ($\sigma$) is applied to the outputs of the main task predicting the probability of each class for an audio clip.
All models reported have an auxiliary branch with two outputs and a softmax activation function. 
A schematic representation of the components and changes w.r.t. the default classification head is shown in \Cref{fig:architecture}. 

% \subsection{Multi-task learning}\label{ss:mtl}
\vspace{-2mm}
\subsection{Loss}\label{ss:mtl}
% Lea et al. could show the benefit of using MTL for stuttering classification by combining 
% a concordence correlation loss function with a focal loss for multi-class classification \cite{lea_SEP28kDatasetStuttering_2021}.

Previous studies have shown the usefulness of multi-task learning (MTL) when training dysfluency detection systems, using either an artificial `any' label, indicating the presence of any dysfluency, or gender classification \cite{lea_SEP28kDatasetStuttering_2021, bayerl_DetectingDysfluenciesStuttering_2022a,sheikh_RobustStutteringDetection_2022}. 

% In our experiments, we use a combination of weighted Binary Cross Entropy (BCE) as the auxiliary loss ($L_{\text{aux}}$), and Focal Loss (FL) as the main task loss ($L_{\text{main}}$) \cite{lin_FocalLossDense_2020}. 

In our experiments, we use a combination of weighted Binary Cross Entropy (BCE) and Focal Loss (FL) \cite{lin_FocalLossDense_2020}. 
FL is an extension of the BCE loss using the $\alpha$ and $\gamma$ parameters to put special emphasis on minority classes to handle class imbalance (see \Cref{tab:distribution}).
% Giving samples from minority classes more weight.

\begin{equation}\label{eq:focal_loss}
    \mathbf{FL}(p_t) = -\alpha (1 - p_t)^\gamma \log{(p_t)}.
    % \vspace{-2mm}
\end{equation}

FL can be used equivalently to BCE loss for multi-label problems by calculating the loss for each class using the output of a given output neuron. 
The total loss is calculated by either summing up the losses for each class or using the mean value as shown in \cref{eq:focal_loss_multi} and used in this work.

\begin{equation}\label{eq:focal_loss_multi}
    \vspace{-2mm}
    \mathbf{FL_{multi}} = \frac{1}{n}\sum^n_1{\mathbf{FL_n}(p_t)}
    % \vspace{-2mm}
\end{equation}

As in \cref{eq:mtlloss}, the final multi-task loss is a weighted sum of $L_{\text{main}}$ and $L_{\text{aux}}$, combining BCE loss for the auxiliary task and FL for the main task of multi-label dysfluency detection. 

\begin{equation}\label{eq:mtlloss}
    \vspace{-2mm}
    \mathbf{L}_{\text{MTL}} = w_{\text{main}} L_{\text{main}} + (1-w_{\text{main}}) L_{\text{aux}}
    \vspace{-2mm}
\end{equation}

\vspace{-4mm}
\section{Experiments}\label{sec:experiments}
\vspace{-2mm}

% Extensive ablation studies are not part of this paper but will be subject of future work. 
Our experiments provide insights into dysfluency detection as a multi-label problem, the influence of training data quantity-, and composition across datasets and languages. 
% the influence of training data quantity, and the transferability and influence of mixing dysfluency training data across datasets and languages. 
Preliminary experiments included modifying the classification head of the W2V2 model by adding an attention mechanism for pooling w.r.t. a trainable token parameter, mean-, or statistical pooling as implemented in \cite{wolf_TransformersStateoftheArtNatural_2020}, or a classification token-based mechanism as used by BERT \cite{devlin_BERTPretrainingDeep_2019}, which led to slightly worse results. 
Using multi-task instead of single-task learning and using FL as a primary loss instead of weighted BCE loss lead to consistently better results, except for word repetitions on the KSoF dataset. 
The models reported here were the best w.r.t. their overall dysfluency detection performance.
% (unweighted average F1).
% Reporting on certain design choices and their influences on detection performance will be subject of future work. 

All experiments were performed using the weights of pre-trained W2V2 base feature extractors. 
The experiments using Multi, Multi-S, \fb, \sepE, and ALL-EN training data, were based on a model that was previously pre-trained on 960h of LibriSpeech \cite{panayotov_LibrispeechASRCorpus_2015} and fine-tuned for ASR (ASR BASE 960h (EN)) \cite{baevski_Wav2vecFrameworkSelfSupervised_2020}.
The experiments using only \ksof data for training were based on a model that was fine-tuned for German ASR using the Common Voice 9.0 dataset (ASR BASE CV-9 (DE)).
\footnote{\protect\url{https://tinyurl.com/3pvj547h}}
Experiments 10 -- 15 utilized weights obtained by training experiment 1 (\Cref{tab:results_data}).
% \footnote{\protect\url{https://huggingface.co/oliverguhr/wav2vec2-base-german-cv9}}. 

The systems were trained using the adamW optimizer \cite{loshchilov_DecoupledWeightDecay_2019}, an initial learning rate of \num{3e-5}, and a batch size of 256 for up to 20 epochs.
The best model was chosen w.r.t. the lowest development loss with early stopping after 5 epochs without improvement. 
During training, the convolutional feature extractor at the beginning of the model was frozen. 
% Early stopping was implemented based on the validation loss. 
The main loss weight $w_{main}$ and the FL parameters $\alpha$ and $\gamma$ were experimentally determined from $w_{main} \in \{0.5, 0.6, ... 0.9\}$,  $\gamma \in \{1, 2, 3\}$, and $\alpha \in \{0.1, 0.2,  \ldots, 0.9 \}$, using grid-search on experiment 1 in \Cref{tab:results_data}.
The best performing parameter configuration was found to be $w_{\text{main}}=0.9$, $\alpha=0.7$, and $\gamma=3$, and used in all subsequent experiments. 
Both, the `any' label and gender classification are equally suited as a MTL target, leading to insignificant performance differences.
The desired regularizing effect is achieved by both. 
\vspace{-2mm}
\section{Results and Discussion}\label{sec:results}
\vspace{-2mm}
% \input{05_results_table}
%%% ONE TABLE EIGHT COLUMNS, RESULTS on EACH DS
\begin{table}[!htb]
\vspace{-3mm}
    \centering
    % \caption{Results for End-to-End multi-label dysfluency detection systems trained and evaluated on different combinations of the training data. Section header indicates the training data used followed by the base-model weights used for fine-tuning. Reporting the F1-score for each dysfluency class in \sepE, \fb ~and KSoF. (
    % \textbf{Mod} = Modified Speech, \textbf{Bl} = Block, \textbf{Int} = Interjection, \textbf{Pro} = Prolongation, \textbf{Snd} = Sound repetition, \textbf{Wd} = Word repetition)}
    \caption{
    Dysfluency detection results (F1-score) for E2E multi-label systems trained and evaluated on different combinations of the training data for each dysfluency class and modifications.
    % in \sepE, \fb, and \ksof 
    (\textbf{Mod} = Modified Speech, \textbf{Bl} = Block, \textbf{Int} = Interjection, \textbf{Pro} = Prolongation, \textbf{Snd} = Sound repetition, \textbf{Wd} = Word repetition).
    Section headers indicate the training data, followed by the W2V2 weights used.
    N/A indicates cases where precision and recall are zero per definition, as there are no labeled clips in the respective split, i.e., F1, is undefined.
    }
    \scalebox{0.90}{
    %Columns are \underlined{Mod}ified, \underlined{Bl}ock, \underlined{Int}erjection, \underlined{Pro}longation, \underlined{S}ou\underlined{nd} rep,m \underlined{W}or\underlined{d} rep.)
\begin{tabular}{c|c|c|c|c|c|c|c}
\toprule
\#& \textbf{Test} & \textbf{Mod}  &  \textbf{Bl} &  \textbf{Int} &  \textbf{Pro} &  \textbf{Snd}  & \textbf{Wd} \\ 
    % \midrule
% 0 & \textbf{SEP28k-E \cite{bayerl_InfluenceDatasetPartitioning_2022}} & - & 0.33 & 0.68 & 0.46 & 0.39 & 0.42 & -      \\
    \midrule
\multicolumn{8}{c}{\textbf{\sepE (ASR BASE 960h EN)}}   \\
    \midrule
1 & \textbf{\sepE} & - & 0.29 & 0.74 & 0.52 & 0.48 & 0.54 \\
2 & FBANK  & - & 0.25 & 0.80 & 0.50 & 0.55 & 0.46 \\
3 & \ksof   & - & 0.33 & 0.61 & 0.32 & 0.43 & \textbf{0.20} \\
    \midrule
\multicolumn{8}{c}{\textbf{\fb (ASR BASE 960h EN)}}   \\
    \midrule
4 & SEP28k & -    & 0.03 & 0.61  &  0.08  & 0.30  & 0.05 \\ 
5 & \textbf{FBANK}  & -    &  0.05 & 0.73  &  0.33 & 0.35 & 0.10  \\
6 & \ksof   & -    & 0.03 &  0.44  &  0.09  & 0.32  & 0.11  \\ 
    \midrule
\multicolumn{8}{c}{\textbf{\ksof (ASR BASE CV-9 DE)}}   \\
    \midrule
7 & \sepE &  N/A   &   0.27 &          0.48 &          0.26 &      0.33 &     0.00  \\
8 & FBANK  &  N/A   &   0.23 &          0.58 &          0.32 &      0.42 &     0.00  \\
9 & \textbf{\ksof}   & \textbf{0.80} &   \textbf{0.61} &          0.76 &          0.48 &      0.42 &     0.00  \\
    \midrule
\multicolumn{8}{c}{\textbf{\fb (SEP-28k-E)}}   \\
    \midrule
10 & \textit{\sepE} &  - & 0.26 & 0.76 & 0.49 & 0.46 & 0.54   \\
11 & \textbf{FBANK}  &  - & \textbf{0.34} & 0.82 & \textbf{0.62} & 0.59 & \textbf{0.49}    \\
12 & \ksof   &  - & 0.38 & 0.58 & 0.37 & 0.40 & 0.16  \\
    \midrule
\multicolumn{8}{c}{\textbf{KSoF (SEP-28k-E)}}   \\
    \midrule
13 & \textit{\sepE} &  N/A  &   0.28 &          0.68 &          0.46 &      0.39 &     0.00    \\
14 & FBANK  &  N/A &   0.26 &          0.65 &          0.53 &      0.50 &     0.00    \\
15 & \textbf{\ksof}   &  0.76  &   0.55 &     \textbf{0.80} &          0.52 &      0.42 &     0.00 \\
    \midrule
    \midrule
\multicolumn{8}{c}{\textbf{ALL-EN (ASR BASE 960h EN)}}   \\
    \midrule
16 & \sepE & -  & \textbf{0.32} & 0.77 &	\textbf{0.54} & \textbf{0.50} & \textbf{0.56} \\
17 & FBANK  & -  & 0.31 & 0.82 &	0.57 & 0.61 & 0.45 \\
18 & \ksof   & -  & 0.41 & 0.57 &	0.34 & 0.42 & 0.14 \\
19 & \textbf{ALL-EN}    & -    & 0.31 & 0.79  &  0.55  & 0.52  & 0.55  \\ 
    \midrule
\multicolumn{8}{c}{\textbf{Multilingual-Small (ASR BASE 960h EN)}}   \\
    \midrule
20 & \sepE & N/A & 0.26 & 0.58 &	0.33 & 0.42 & 0.22 \\
21 & FBANK & N/A  & 0.27 & 0.69 &	0.49 & 0.55 & 0.30 \\
22 & \ksof & 0.74  & 0.49 & 0.67 &	0.46 & 0.44 & 0.06 \\
23 & Multi-S & 0.75  & 0.46 & 0.68 & 0.49 & 0.42  & 0.21   \\ 
    \midrule
\multicolumn{8}{c}{\textbf{Multilingual (ASR BASE 960h EN)}}   \\
    \midrule
24 & \sepE & N/A   & 0.31 & \textbf{0.79}  &  \textbf{0.54}  & \textbf{0.50}  & \textbf{0.56} \\ 
25 & FBANK  & N/A   & 0.31 & \textbf{0.83}  &  0.57  & \textbf{0.63}  & 0.48  \\ 
26 & \ksof   & 0.79   & 0.52 & 0.78  &  \textbf{0.55}  & \textbf{0.46}  & 0.08  \\ 
27 & \textbf{Multi}    & 0.79 & 0.36 & 0.79  &  0.55  & 0.51  & 0.53  \\ 
\bottomrule
\end{tabular}
}
    \label{tab:results_data}

\vspace{-6mm}
\end{table}

In the interest of brevity, we only report F1-scores for all dysfluency types and modifications in \Cref{tab:results_data}.
Results were balanced w.r.t. \textit{precision} and \textit{recall}.

\textbf{Bl} can be difficult to detect using only one modality.
Clinicians rely on signs of physical tension and grasping for air when assessing blocks. 
This is also reflected by low inter-rater reliability (IRR)(Fleiss $\kappa$, 0.25 SEP-28k, 0.37 \ksof) for \textbf{Bl} reported for the datasets \cite{lea_SEP28kDatasetStuttering_2021, bayerl_KSoFKasselState_2022}. 
% In this light detection results must be taken with a grain of salt.
We hypothesize that the consistently better results for \textbf{Bl} on \ksof (exp.~3,~9,~12,~15,~18,~22,~26) might be due to \ksof consisting of therapy recordings that mostly include PWS who, on average, have more pronounced symptoms. 
% as opposed to people appearing on a podcast. 
The slightly higher $\kappa$ supports this observation.  

\textbf{Wd} are easy to detect for humans, who assess the meaning of a sentence while hearing it, which a system relying on acoustics features alone cannot deliver. 
Acoustically, there is little to no difference between words and their immediate repetition.
There is an obvious language component to detecting \textbf{Wd}, as evidenced by the ALL-EN and multi-lingual experiments that perform best on \sepE but fail to recognize German \textbf{Wd} (exp.~16,~18,~24,~26).
Data quantity also plays a role, as evidenced by the failure to detect word repetitions on models trained using only \ksof or \fb or their combination Multi-S (exp.~4-9,~20-23). 
\ksof and \fb contain only 212 and 430 clips labeled as word repetitions (comp.~\Cref{tab:distribution}) which is not enough to learn such an acoustically diverse pattern with this method. 

\textbf{Int}s are generally well detected by all systems and across all datasets. 
% The IRR of the labels indicates, that there is little room for actual improvement on this dysfluency type without overfitting to the datasets (0.57 SEP-28k, 0.78 \ksof) \cite{lea_SEP28kDatasetStuttering_2021,bayerl_KSoFKasselState_2022}. 
The best results for \ksof were achieved with the model pre-trained on English stuttering data and fine-tuned on German data (exp.~15). 
The detection results for \fb and \sepE profit from adding German training data (exp.~24,~25), as function and acoustic composition of those functional dysfluencies are similar between English and German \cite[p.17]{belz_PhonetikAhUnd_2021}.
The jointly learned system creates features that transfer well between languages.
Adding more training data introduces information that helps the model generalize. 
% TODO rewrite last sentence

%  are detected reliably in all models trained using the \ksof data. 
% As an universally taught acquired pattern they are relatively easy to detect from acoustics. 
\textbf{Mod}s are characterized by prolonged uninterrupted phonation and soft voice onset, making this universally taught pattern relatively easy to detect acoustically \cite{bayerl_KSoFKasselState_2022}.  
The model based on German ASR and fine-tuned using \ksof achieved the highest F1 on \ksof (exp.~9). 
Evaluating the model on \sepE and \fb leads to quite a few false positives, as the KSoF\=/only trained model does not generalize very well.
This indicates that the model learned to recognize \textbf{Mod} but has not seen enough other dysfluency data to differentiate outside its own training domain.  
Evaluating the recognition of \textbf{Mod} with the Multilingual model shows that language plays a role, as all samples of modified speech are in German. 
Adding more dysfluency training data helps the model to learn differentiate modified speech from other dysfluency patterns better, as indicated by the improved F1 score for \textbf{Mod} from Multi-S to Multi (exp.~22,~23,~26,~27).
At the same time the amount of \textbf{Mod} clips stay fixed. 

% The multilingual model performed slightly worse on \ksof and Multilingual (exp.~33,~35) than the model based on pre-trained German ASR, showing that only very few misclassifications happen outside of \ksof when using the Multilingual model. .
% Language and recording conditions seem to play a role in the classification results for the multilingual model. 
% probably detects the language in the clip and not only the modified pattern. 
% This is supported by the fact that all misclassifications happen to be other clips from \ksof.
% b as all examples labeled as modified speech are from the German \ksof dataset and a large portion of recognizing those clips correctly might be due to the difference in language. 

% 13/15 -> lots misclassifications towards other dsyfluencies, model only trained on ksof, not very well generalizability towards the english stuttering data (see othe results)
% 23/25 -> lots misclassifications towards other dsyfluencies, but much less than model without sep28k weights, better generalization towards dysfluencies in gerneral, even though f1 was less (could that indicate there actually is fluency shaping in sep28k?

% ksof on modifications best in fine-tuned: knowledge about other stutters helps to get better results,

\textbf{Pro} and \textbf{Snd} change the natural flow and rhythm of speech independent of the language.
Acoustically, repeating a plosive in English is similar to German, the same holds true for the prolongation of, e.g, a vowel. 
Differences mostly stem from the respective phoneme inventories. 
Therefore both patterns profit from more training data and transfer learning.  
The best results for these dysfluency types were achieved using the model trained on Multi (exp.~24-26). 
The model trained on Multi profits from jointly learning these patterns for both languages. 
Only \textbf{Pro} for \fb (exp. 25) perform worse than the system pre-trained on \sepE and fine-tuned on \fb (exp. 16).

Comparing results for \sepE and \fb using the ALL-EN and Multilingual model shows that the overall performance does not suffer from adding the German stuttering data. 
The model even slightly profits from the additional training data (exp.~16,~17,~24,~25).
At the same time, results achieved on \ksof improved substantially (exp.~18,~26). 
Adding the 3500 clips from the \ksof training set to the joint training data enabled the model to generalize to another language and learn to classify an additional pattern. 

% mix/eng results evidence of little misclassification (.79) as amount of events did not increase/decrease / probbly also due to language

% From the results of the models trained solely on \fb and \ksof the statement can be made, that a W2V2 based model trained on multiple dysfluency types needs more 
% than data to generalize. (fb train set: \~2500 even smaller than ksof\~3500)

\vspace{-2mm}
\section{Conclusion}\label{sec:conclusion}
\vspace{-2mm}

This paper has introduced a W2V2-based End-2-End stuttering detection and classification system able to detect and classify five types of stuttering, and speech modifications, achieving state-of-the-art dysfluency detection results on \sepE.
The best results achieved on \ksof and \fb are comparable to previous expert systems that only detect a single kind of dysfluency \cite{bayerl_DetectingDysfluenciesStuttering_2022a}. 

In future work, we will explore data augmentation and contrastive training, as data quantity and diversity play a role in creating more robust dysfluency detection systems.
Especially with \textbf{Wd} and \textbf{Bl}, a system relying only on the interpretation of acoustic features reaches its limits. 
While there might be no obvious solutions to improve actual recognition results for \textbf{Bl}, with the limited reliability of the labels, \textbf{Wd} recognition might profit from integrating phoneme posterior based features \cite{klumpp_PhoneticFootprintParkinson_2022} or other ASR-based features \cite{bayerl_AutomatedAssessmentStuttering_2020}. 

% An increase in the amount of training data, by using data augmentation techniques such as reverberation and speed perturbation. 
% This work could clearly show that multiple datasets could be used in joint training and achieve good results on each dataset without losing predictive power on the other datasets. 
% Another possible direction is to improve results mainly on word-repetitions that have highly reliable labels but are hard to catch from acoustics only. 
% Integrating some pre-processing steps including phoneme or speech recognition could help with taht. 
% A combination of an expert system in conjunction with the existing W2V2-based end-2-end system could leverage expert knowledge about different kinds of stutter and finding appropriate feature extraction methods to suit them.

\FloatBarrier
\newpage

\vfill\pagebreak

\bibliographystyle{IEEEbib}
\footnotesize{\bibliography{zotero.bib}}

\begin{thebibliography}{10}

\bibitem{wingate_FluencyDisfluencyDysfluency_1984}
Marcel~E. Wingate,
\newblock ``Fluency, disfluency, dysfluency, and stuttering,''
\newblock {\em Journal of Fluency Disorders}, vol. 9, no. 2, pp. 163--168, May
  1984.

\bibitem{lickley_DisfluencyTypicalStuttered_2017}
Robin Lickley,
\newblock ``Disfluency in typical and stuttered speech,''
\newblock {\em Fattori sociali e biologici nella variazione fonetica}, , no. 3,
  pp. 373, 2017.

\bibitem{bayerl_DetectingDysfluenciesStuttering_2022a}
Sebastian~P. Bayerl, Dominik Wagner, Elmar Noeth, and Korbinian Riedhammer,
\newblock ``Detecting {{Dysfluencies}} in {{Stuttering Therapy Using}} wav2vec
  2.0,''
\newblock in {\em Interspeech 2022}. Sept. 2022, pp. 2868--2872, {ISCA}.

\bibitem{harvill_FramelevelStutterDetection_2022}
John Harvill, Mark {Hasegawa-Johnson}, and Chang~D. Yoo,
\newblock ``Frame-level stutter detection,''
\newblock in {\em Proc. {{Interspeech}} 2022}, 2022, pp. 2843--2847.

\bibitem{kourkounakis_DetectingMultipleSpeech_2020}
Tedd Kourkounakis, Amirhossein Hajavi, and Ali Etemad,
\newblock ``Detecting {{Multiple Speech Disfluencies}} using a {{Deep Residual
  Network}} with {{Bidirectional Long Short-Term Memory}},''
\newblock in {\em {{ICASSP}} 2020-2020 {{IEEE International Conference}} on
  {{Acoustics}}, {{Speech}} and {{Signal Processing}} ({{ICASSP}})}. 2020, pp.
  6089--6093, {IEEE}.

\bibitem{grosz_Wav2vec2basedParalinguisticSystems_2022}
Tam{\'a}s Gr{\'o}sz, Dejan Porjazovski, Yaroslav Getman, Sudarsana Kadiri, and
  Mikko Kurimo,
\newblock ``Wav2vec2-based paralinguistic systems to recognise vocalised
  emotions and stuttering,''
\newblock in {\em Proceedings of the 30th {{ACM}} International Conference on
  Multimedia}, {New York, NY, USA}, 2022, {{MM}} '22, pp. 7026--7029,
  {Association for Computing Machinery}.

\bibitem{sheikh_EndtoEndSelfSupervisedLearning_2022}
Shakeel~A. Sheikh, Md~Sahidullah, Slim Ouni, and Fabrice Hirsch,
\newblock ``End-to-end and self-supervised learning for {{ComParE}} 2022
  stuttering sub-challenge,''
\newblock in {\em Proceedings of the 30th {{ACM}} International Conference on
  Multimedia}, {New York, NY, USA}, 2022, {{MM}} '22, pp. 7104--7108,
  {Association for Computing Machinery}.

\bibitem{montacie_AudioFeaturesWav2Vec_2022}
Claude Montaci{\'e}, Marie-Jos{\'e} Caraty, and Nikola Lackovic,
\newblock ``Audio features from the {{Wav2Vec}} 2.0 embeddings for the {{ACM}}
  multimedia 2022 stuttering challenge,''
\newblock in {\em Proceedings of the 30th {{ACM}} International Conference on
  Multimedia}, {New York, NY, USA}, 2022, {{MM}} '22, pp. 7195--7199,
  {Association for Computing Machinery}.

\bibitem{you_MaskedModelingbasedAudio_2022}
Kang You, Kele Xu, Boqing Zhu, Ming Feng, Dawei Feng, Bo~Liu, Tian Gao, and
  Bo~Ding,
\newblock ``Masked modeling-based audio representation for {{ACM}} multimedia
  2022 computational paralinguistics {{ChallengE}},''
\newblock in {\em Proceedings of the 30th {{ACM}} International Conference on
  Multimedia}, {New York, NY, USA}, 2022, {{MM}} '22, pp. 7060--7064,
  {Association for Computing Machinery}.

\bibitem{eyben_OpensmileMunichVersatile_2010}
Florian Eyben, Martin W{\"o}llmer, and Bj{\"o}rn Schuller,
\newblock ``Opensmile: The munich versatile and fast open-source audio feature
  extractor,''
\newblock in {\em Proceedings of the International Conference on {{Multimedia}}
  - {{MM}} '10}, {Firenze, Italy}, 2010, p. 1459, {ACM Press}.

\bibitem{howell_UniversityCollegeLondon_2009}
Peter Howell, Stephen Davis, and Jon Bartrip,
\newblock ``The {{University College London Archive}} of {{Stuttered Speech}}
  ({{UCLASS}}),''
\newblock {\em Journal of Speech, Language, and Hearing Research}, vol. 52, no.
  2, pp. 556--569, Apr. 2009.

\bibitem{lea_SEP28kDatasetStuttering_2021}
Colin Lea, Vikramjit Mitra, Aparna Joshi, Sachin Kajarekar, and Jeffrey~P.
  Bigham,
\newblock ``{{SEP-28k}}: {{A Dataset}} for {{Stuttering Event Detection}} from
  {{Podcasts}} with {{People Who Stutter}},''
\newblock in {\em {{ICASSP}} 2021 - 2021 {{IEEE International Conference}} on
  {{Acoustics}}, {{Speech}} and {{Signal Processing}} ({{ICASSP}})}, {Toronto,
  ON, Canada}, June 2021, pp. 6798--6802, {IEEE}.

\bibitem{sheikh_RobustStutteringDetection_2022}
Shakeel~A Sheikh, Fabrice Hirsch, and Slim Ouni,
\newblock ``Robust {{Stuttering Detection}} via {{Multi-task}} and
  {{Adversarial Learning}},''
\newblock in {\em 2022 30th {{European Signal Processing Conference}}
  ({{EUSIPCO}})}, 2022, p.~5.

\bibitem{mohapatra_SpeechDisfluencyDetection_2022}
Payal Mohapatra, Akash Pandey, Bashima Islam, and Qi~Zhu,
\newblock ``Speech disfluency detection with contextual representation and data
  distillation,''
\newblock in {\em Proceedings of the 1st {{ACM}} International Workshop on
  Intelligent Acoustic Systems and Applications}, {New York, NY, USA}, 2022,
  {{IASA}} '22, pp. 19--24, {Association for Computing Machinery}.

\bibitem{jouaiti_DysfluencyClassificationStuttered_2022}
Melanie Jouaiti and Kerstin Dautenhahn,
\newblock ``Dysfluency {{Classification}} in {{Stuttered Speech Using Deep
  Learning}} for {{Real-Time Applications}},''
\newblock in {\em {{ICASSP}} 2022 - 2022 {{IEEE International Conference}} on
  {{Acoustics}}, {{Speech}} and {{Signal Processing}} ({{ICASSP}})},
  {Singapore, Singapore}, May 2022, pp. 6482--6486, {IEEE}.

\bibitem{bayerl_InfluenceDatasetPartitioning_2022}
Sebastian~P. Bayerl, Dominik Wagner, Elmar N{\"o}th, Tobias Bocklet, and
  Korbinian Riedhammer,
\newblock ``The {{Influence}} of {{Dataset Partitioning}} on {{Dysfluency
  Detection Systems}},''
\newblock in {\em Text, {{Speech}}, and {{Dialogue}}}, Petr Sojka, Ivan Kope{\v
  c}ek, Karel Pala, and Ale{\v s} Hor{\'a}k, Eds. {Springer International
  Publishing}, 2022.

\bibitem{bayerl_KSoFKasselState_2022}
Sebastian~P. Bayerl, Alexander {Wolff von Gudenberg}, Florian H{\"o}nig, Elmar
  Noeth, and Korbinian Riedhammer,
\newblock ``{{KSoF}}: {{The Kassel State}} of {{Fluency Dataset}} -- {{A
  Therapy Centered Dataset}} of {{Stuttering}},''
\newblock in {\em Proceedings of the Language Resources and Evaluation
  Conference}, {Marseille, France}, June 2022, pp. 1780--1787, {European
  Language Resources Association}.

\bibitem{bernsteinratner_FluencyBankNew_2018}
Nan Bernstein~Ratner and Brian MacWhinney,
\newblock ``Fluency {{Bank}}: {{A}} new resource for fluency research and
  practice,''
\newblock {\em Journal of Fluency Disorders}, vol. 56, pp. 69--80, June 2018.

\bibitem{baevski_Wav2vecFrameworkSelfSupervised_2020}
Alexei Baevski, Yuhao Zhou, Abdelrahman Mohamed, and Michael Auli,
\newblock ``Wav2vec 2.0: {{A Framework}} for {{Self-Supervised Learning}} of
  {{Speech Representations}},''
\newblock in {\em Advances in {{Neural Information Processing Systems}}},
  H.~Larochelle, M.~Ranzato, R.~Hadsell, M.~F. Balcan, and H.~Lin, Eds. 2020,
  vol.~33, pp. 12449--12460, {Curran Associates, Inc.}

\bibitem{xu_ExploreWav2vecMispronunciation_2021}
Xiaoshuo Xu, Yueteng Kang, Songjun Cao, Binghuai Lin, and Long Ma,
\newblock ``Explore wav2vec 2.0 for {{Mispronunciation Detection}},''
\newblock in {\em Interspeech 2021}. Aug. 2021, pp. 4428--4432, {ISCA}.

\bibitem{wolf_TransformersStateoftheArtNatural_2020}
Thomas Wolf, Lysandre Debut, Victor Sanh, Julien Chaumond, Clement Delangue,
  Anthony Moi, Pierric Cistac, Tim Rault, R{\'e}mi Louf, Morgan Funtowicz, Joe
  Davison, Sam Shleifer, Patrick von Platen, Clara Ma, Yacine Jernite, Julien
  Plu, Canwen Xu, Teven~Le Scao, Sylvain Gugger, Mariama Drame, Quentin Lhoest,
  and Alexander~M. Rush,
\newblock ``Transformers: {{State-of-the-Art Natural Language Processing}},''
\newblock in {\em Proceedings of the 2020 {{Conference}} on {{Empirical
  Methods}} in {{Natural Language Processing}}: {{System Demonstrations}}},
  {Online}, Oct. 2020, pp. 38--45, {Association for Computational Linguistics}.

\bibitem{lin_FocalLossDense_2020}
Tsung-Yi Lin, Priya Goyal, Ross Girshick, Kaiming He, and Piotr Dollar,
\newblock ``Focal {{Loss}} for {{Dense Object Detection}},''
\newblock {\em IEEE Transactions on Pattern Analysis and Machine Intelligence},
  vol. 42, no. 2, pp. 318--327, Feb. 2020.

\bibitem{devlin_BERTPretrainingDeep_2019}
Jacob Devlin, Ming-Wei Chang, Kenton Lee, and Kristina Toutanova,
\newblock ``{{BERT}}: {{Pre-training}} of {{Deep Bidirectional Transformers}}
  for {{Language Understanding}},''
\newblock {\em arXiv:1810.04805 [cs]}, May 2019.

\bibitem{panayotov_LibrispeechASRCorpus_2015}
Vassil Panayotov, Guoguo Chen, Daniel Povey, and Sanjeev Khudanpur,
\newblock ``Librispeech: {{An ASR}} corpus based on public domain audio
  books,''
\newblock in {\em 2015 {{IEEE International Conference}} on {{Acoustics}},
  {{Speech}} and {{Signal Processing}} ({{ICASSP}})}, {South Brisbane,
  Queensland, Australia}, Apr. 2015, pp. 5206--5210, {IEEE}.

\bibitem{loshchilov_DecoupledWeightDecay_2019}
Ilya Loshchilov and Frank Hutter,
\newblock ``Decoupled {{Weight Decay Regularization}},''
\newblock {\em arXiv:1711.05101 [cs, math]}, Jan. 2019.

\bibitem{belz_PhonetikAhUnd_2021}
Malte Belz,
\newblock {\em {Die Phonetik von \"ah und \"ahm: Akustische Variation von
  F\"ullpartikeln im Deutschen}},
\newblock {Springer Berlin Heidelberg}, {Berlin, Heidelberg}, 2021.

\bibitem{klumpp_PhoneticFootprintParkinson_2022}
Philipp Klumpp, Tom{\'a}s {Arias-Vergara}, Juan~Camilo {V{\'a}squez-Correa},
  Paula~Andrea {P{\'e}rez-Toro}, Juan~Rafael {Orozco-Arroyave}, Anton Batliner,
  and Elmar N{\"o}th,
\newblock ``The {{Phonetic Footprint}} of {{Parkinson}}'s {{Disease}},''
\newblock {\em Computer Speech \& Language}, vol. 72, pp. 101321, Mar. 2022.

\bibitem{bayerl_AutomatedAssessmentStuttering_2020}
Sebastian~P. Bayerl, Florian H{\"o}nig, Jo{\"e}lle Reister, and Korbinian
  Riedhammer,
\newblock ``Towards {{Automated Assessment}} of {{Stuttering}} and {{Stuttering
  Therapy}},''
\newblock in {\em Text, {{Speech}}, and {{Dialogue}}}, Petr Sojka, Ivan Kope{\v
  c}ek, Karel Pala, and Ale{\v s} Hor{\'a}k, Eds., {Cham}, 2020, vol. 12284,
  pp. 386--396, {Springer International Publishing}.

\end{thebibliography}

\end{document}